\documentclass[11pt, executivepaper]{article}
\usepackage[utf8]{inputenc}
\usepackage[T1]{fontenc}
\usepackage{natbib}
\usepackage{amsmath}
\usepackage{mathtools}
\usepackage{xcolor}
\usepackage{amsfonts}
\usepackage{graphicx}
\usepackage{enumitem}
\usepackage{geometry}
\usepackage{hyperref}
\hypersetup{colorlinks= true, allcolors=blue}
\setcitestyle{aysep={}}
\begin{document}

\title{\textbf{Beables, Primitive Ontology and Beyond: How Theories Meet the World}}

\author{Andrea Oldofredi\thanks{Contact Information: Universit\'e de Lausanne, Section de Philosophie, 1015 Lausanne, Switzerland. E-mail: Andrea.Oldofredi@unil.ch}}

\maketitle

\begin{abstract}
Bohm and Bell's approaches to the foundations of quantum mechanics share notable features with the contemporary Primitive Ontology perspective and Esfeld and Deckert minimalist ontology. For instance, all these programs consider ontological clarity a necessary condition to be met by every theoretical framework, promote scientific realism also in the quantum domain and strengthen the explanatory power of quantum theory. However, these approaches remarkably diverge from one another, since they employ different metaphysical principles leading to conflicting \emph{Weltanschaaungen}.\ The principal aim of this essay is to spell out the relations as well as the main differences existing among such programs, which unfortunately remain often unnoticed in literature. Indeed, it is not uncommon to see Bell's views conflated with the PO programme, and the latter with Esfeld and Deckert's proposal. It will be our task to clear up this confusion.
\vspace{5mm}

\end{abstract}
\vspace{5mm}
\begin{quote}
\center \emph{To appear in the volume ``Quantum Mechanics and Fundamentality: Naturalizing Quantum Theory between Scientific Realism and Ontological Indeterminacy'', edited by Valia Allori, Springer Nature.}
\end{quote}
\clearpage 

\tableofcontents
\vspace{5mm}

\section{Introduction}

Quantum Mechanics (QM) is one of the most efficient descriptions of the inherent structure of matter and its behavior.\ As is known, however, this theory represents elementary objects in a way that drastically changed our conception of reality, speaking about intrinsically indeterminate systems, non-local interactions, wave-particle dualities, \emph{etc.}. Indeed, it is still a pivotal philosophical challenge to understand what scientific image of the world it provides---i.e.\ to comprehend its ontology, and how the macroscopic realm can be explained from it. Referring to this, it is worth noting that quantum theory is affected by several conceptual and technical conundra; hence, we cannot consider it the final word about ontological matters, notwithstanding its empirical success. In particular, scholars agree that QM contains physically ill-defined notions within its axioms, and inconsistencies among its dynamical laws---resulting in the (in)famous quantum measurement problem.

Against this background, the Primitive Ontology (PO) approach has been advanced to overcome these issues. Today, in fact, we have several PO theories that resolve the quantum puzzles and that can be considered serious alternatives to the standard formulation of QM, as for instance Bohmian Mechanics (BM), dynamical reduction models as the Ghirardi-Rimini-Weber (GRW) theories, \emph{etc.}.\ Although the expression ``primitive ontology'' has been introduced in \cite{Durr:1992aa}, the roots of this programme can be found in Bohm and Bell's foundational works.\ 
Following their steps, the modern formulation of the PO perspective requires that well-defined physical theories explain the macroscopic regime through a more fundamental, microscopic picture, postulating a clear primitive ontology---i.e.\ specifying a set of theoretical entities representing real objects moving in physical space---and a consistent dynamics.\ Moreover, elaborating on the notion of primitive ontology, \cite{Esfeld:2017} extended and modified this concept proposing an atomistic ontology that is taken to be valid at every energy/length scale between the classical regime and the quantum field theoretical level. 

Bohm and Bell's approaches share notable features with the modern PO perspectives; for instance, they consider ontological clarity a necessary condition to be met by every theoretical framework, promote scientific realism also in the quantum domain and strengthen the explanatory power of quantum theory.\footnote{I assume that the reader has some familiarity with this literature.}\ However, these positions remarkably diverge from one another, since they employ different metaphysical principles leading to conflicting \emph{Weltanschaaungen}. The principal aim of this essay is to spell out the relations and the main differences existing among such programs, which unfortunately remain often unnoticed in literature. Indeed, it is not uncommon to see Bell's methodological approach conflated with the PO programme, and the latter with Esfeld and Deckert's views. It will be our task to clear up this confusion.

The paper is organized as follows:\ section \ref{MI} analyzes Bohm's reflections on the nature of theoretical entities and physical laws, taking into account also his metaphysical infinitism. Section \ref{Bell} focuses on Bell's theory of local beables.\ In section \ref{PO} the modern formulation of the PO programme is presented underlying the main metaphysical and methodological differences w.r.t.\ Bohm and Bell's approaches. Section \ref{ED} introduces Esfeld and Deckert's minimalist ontology and illustrates how this perspective diverges under important respects with all the previous proposals. Finally, section \ref{conc} concludes the paper.

\section{David Bohm, Pluralism and Infinitism}
\label{MI}

Ontological clarity is an essential feature of physical theories according to Bohm. On his view the problems of the standard interpretation of QM come from its obscure metaphysical content.\ Bohm was deeply unsatisfied by the lack of explanation of measurement results, since quantum theory excludes a precise characterization of individual systems as well as an accurate description of their dynamical evolution ``without proving that such a renunciation is necessary'' (\cite{Bohm:1952}:168).\ Furthermore, he argued that the empirical robustness of QM and its contingent mathematical structure are not sufficient to exclude a priori other ontologically clearer formulations.  

To tame these issues, Bohm joined the foundational debate proposing an alternative interpretation of QM where quantum particles have definite positions and velocities, and are dynamically guided by a real $\psi$-field, challenging the metaphysical indeterminacy of the standard formulation.\ 
Hence, postulating a dual ontology of particles and fields, he showed that an individual description of quantum systems following continuous trajectories in space was not only mathematically possible, but also physically consistent.\ Moreover, this theory is able to explain measurement outcomes without invoking mysterious collapses of $\psi$, strengthening the explanatory power of quantum theory. In addition, Bohm extended his approach to electromagnetism employing  fields---different from $\psi$---as fundamental entities, showing that he was open to implement various ontologies in different theories (cf.\ Appendix A of \cite{Bohm:1952}). The common trait between such theories is their metaphysical clarity, since Bohm always specified which variables represent matter and how they dynamically behave.

In addition, he claimed that QM---as any other theory---has a limited validity and that ``at distances of the order of 10$^{-13}$cm or smaller and for times of the order of this distance divided by the velocity of light or smaller, present theories become so inadequate that it is generally believed that they are probably not applicable'' (\cite{Bohm:1952}, footnote 6). At these regimes he expected that new ontologies and new theories will be discovered.\ More generally, Bohm thought that physical theories have limited domains of  application at precise length/energy scales, and remarked that what is measured at a particular level---thereby what exists at that scale---depends on the theory at hand:
\begin{quote}
our epistemology is determined to a large extent by the existing theory. It is therefore not wise to specify the possible forms of future theories in terms of purely epistemological limitations deduced from existing theories (\cite{Bohm:1952}:188).
\end{quote}

In his view this is a direct objection against the empiricist basis of QM, which follows the positivistic principle of not accepting entities that cannot be currently observed. According to Bohm, this is a poor working hypothesis, since the history of physics showed the fruitfulness to assume the existence of certain items before their empirical discovery, and the atomic theory is a pivotal example.\ On the other hand, he rejected another perspective, later called ``mechanistic philosophy'', for which reality can be fully explained starting from a fixed set of entities, and a restricted set of laws---something close to what philosophers call foundationalism.\footnote{In section \ref{ED} we will see that \cite{Esfeld:2017} proposed a view similar to the mechanistic philosophy described by Bohm. It should be noted, however, that according to these authors physical laws can change in time in order to accomodate the discovery of new phenomena.} He warns us not to expect such a knowledge ``because there are almost certainly more elements in existence than we possibly can be aware of at any particular stage of scientific development.\ Any specified element, however, can in principle ultimately be discovered, but never all of them'' (\cite{Bohm:1952}:189). This is an hint of the metaphysical infinitism endorsed in \emph{Causality and Chance in Modern Physics} to which we now turn.

In \cite{Bohm:1957} we find detailed objections against mechanistic philosophy.\ Firstly, history of physics disconfirms the basic tenets of this view, since the revolutions that occurred from Newton to this day radically changed the entities and the laws of our theories.\ Moreover, future frameworks will be as revolutionary as QM was w.r.t.\ classical physics.\ Secondly, the assumptions concerning the final character of any particular ontology are neither necessary, nor provable, because future theories may demonstrate its limited validity.\footnote{Referring to this Bohm writes: ``Newton's laws of motion, regarded as absolute and final for over two hundred years, were eventually found to have a limited domain of validity, these limits having finally been expressed with the aid of the quantum theory and the theory of relativity'' (\cite{Bohm:1957}:90).}\ Finally, foundationalism contravenes the scientific method, since the latter imposes that every object and law must be continuously subjected to verification. This process of testing may end up in contradiction with new discoveries or new domains of science. Looking at how physics evolved, says Bohm, such contradictions not only systematically appeared, but also led to a deeper comprehension of the world. 
 
Contrary to such view, he proposes a form of infinitism.\ In essence, Bohm stated that physical sciences and experimental data push us to a conception of nature composed by an infinity of different entities, which do not depend ontologically on a fixed set of absolutely fundamental objects (\cite{Bohm:1957}: 91).\ According to this view of science, physical theories do not always lead us closer to a fundamental ground, but instead show the infinite complexity of nature. Furthermore, he believes that empirical data cannot a priori provide any justification to metaphysical restrictions concerning a particular set of entities to be chosen as absolutely ontologically independent.\ On the contrary, conforming to Bohm's infinitism, scientific practice always discloses new entities, laws and phenomena which contribute to our continuous, never-ending process of understanding the structure of reality. 

However, although Bohm denied the existence of a fundamental level, he firmly believed that every theory must be ontologically unambiguous in its domain of application.\ Therefore, theoretical frameworks must provide a clear ontology to be applied at the relevant energy/length scale, meaning that the entities forming the basic ontology of a given theory can be considered \emph{relatively} fundamental. Referring to this, he stated that 
\begin{quote}
[a]ny given set of qualities and properties of matter and categories of laws that are expressed in terms of these qualities and properties is in general applicable only within limited contexts, over limited ranges of conditions and to limited degrees of approximation, these limits being subject to better and better determination with the aid of further scientific research (\cite{Bohm:1957}:91).
\end{quote}

\noindent Thus, Bohm claimed that a well-defined physical theory should provide a clear ontological picture for the domain in which it is a reliable description of physical phenomena. Nonetheless, its ontology can be substantially modified with the progress of scientific research.\ This certainly exemplifies Bohm's scientific pluralism and his heterodox metaphysical views w.r.t.\ to the dominant paradigm towards the interpretation of QM (cf.\ \cite{vanStrien:2019}). 

As we can see, the seeds of the PO programme---i.e.\ the requirements of ontological clarity and explanatory robustness---have been sowed.\ They will germinate in Bell's theory of local beables which we are going to discuss.

\section{Local and Non-local Beables}
\label{Bell}

It is well-known that Bell was disappointed by the lack of ontological clarity in QM, which in his opinion was the source of its inconsistency and redundancy. Inconsistency because the dynamical laws of the theory contradict each other\footnote{These laws are the Schr\"odinger equation and the collapse postulate.}; redundancy because some central notions contained in its axioms, e.g.\ \emph{observation} and \emph{observable}, should be derived from more fundamental concepts, the beables of the theory:
\begin{quote}
[t]he concept of `observable' lends itself to very precise \emph{mathematics} when identified with `self-adjoint operator'. But physically, it is a rather woolly concept. It is not easy to identify precisely which physical processes are to be given the status of `observations' and which are to be relegated to the limbo between one observation and another. So it could be hoped that some increase in precision might be possible by concentration on the \emph{be}ables, which can be described in `classical terms', because they are there (\cite{Bell:2004aa}:52).
\end{quote}

\noindent According to Bell, a well-defined physical theory $T$ must postulate a clear ontology, or in his jargon, a set of local beables. These are the theoretical entities of $T$ referring to real objects ascribed to bounded regions of space-time, i.e.\ they correspond to the elements of reality that the theory postulates to exist independently on any observation.\ It is worth noting that the beables cannot be derived from other more fundamental notions of $T$; hence, they establish what is fundamental \emph{in the context of T}.

Moreover, together with dynamical laws\footnote{Remarkably, according to Bell in the context of a given theory $T$ there is a distinction between physical and non-physical entities:\ the former denote the beables of the theory, the latter are the mathematical structures needed to formulate dynamical laws for the variables representing matter in space; cf.\ \cite{Bell:2004aa}, Chapter 7.}, the beables explain the physical phenomena lying within the domain of application of the theory under consideration, connecting its formal structure to the macroscopic world accessible to us:
\begin{quote}
[t]he beables must include the settings of switches and knobs on experimental equipment, the currents in coils, and the reading of instruments. `Observables' must be \emph{made}, somehow, out of beables.\ The theory of local beables should contain, and give precise physical meaning to, the algebra of local observables (\emph{ibid}.).
\end{quote}

But how do we select a particular set of beables for a theory?\ In answering this question Bell's pluralist attitude becomes manifest, since he claimed that the only necessary requirements for an ontology is to state explicitly which variables represent matter, and to provide unambiguous explanation of macroscopic observations.\footnote{To this regard Bell writes that ``what is essential is to be able to define the positions of things, including the positions of pointers or (the modern equivalent) of ink of computer output'' (\cite{Bell:2004aa}:175).\ Cf.\ also \cite{Bell:2004aa}, Chapter 5.}

These liberal requirements do not impose any strict limitation to the selection of a certain ontology.\ Such a freedom of choice is justified because (i) there are always several adequate options to account for the macroscopic data that a certain theory must explain, (ii) physical theories have a provisional character.\ Therefore, according to Bell, there is no one-to-one relation between the beables of a theory and the physical world, meaning that we cannot definitively establish whether a certain ontology is the correct, ultimate description of reality.\ To this regard, Bell used ``the term `beable' rather than some more committed term like `being' or `beer' to recall the essentially tentative nature of any physical theory. Such a theory is at best a \emph{candidate} for the description of nature. Terms like `being', `beer', `existent', etc., would seem to me lacking in humility. In fact `beable' is short for `maybe-able' '' (\cite{Bell:2004aa}:174).

Indeed, there are various examples of beables implemented in different classical and quantum theories, as for instance electromagnetic fields, point particles, matter density fields, strings \emph{etc.}.\footnote{\cite{Maudlin:2016} underlines that a similar choice is available also for the specification of the space-time structure of a given theory.} Bell himself, although he was a supporter of the pilot-wave theory (\cite{Bell:2004aa}, Chapter 17), proposed a rival picture for non-relativistic QM, namely a GRW theory with a flash ontology (\emph{ibid.}, Chapter 22). Moreover, he extended Bohm's approach to Quantum Field Theory (QFT) implementing an ontology of fermion number density (\emph{ibid.}, Chapter 19). Such beables have no classical analogues, implying an ontological discontinuity between the classical and the quantum regime; however, such a discontinuity was tolerated by him.\ Similarly, Bell's approach allows that a single theory can postulate the existence of different kinds of objects, i.e.\ it can provide a multi-category ontology, as for instance in classical electromagnetism, where we find both point particles and fields.\ Thus, the theory of local beables does not require that reality must be reduced to a unique category of physical entities. These facts certainly denote his liberal views about the ontology of physical theories, in close analogy with Bohm's pluralist perspective. 
Finally, it is interesting to note the Bell admitted the existence of \emph{non-local} beables.\footnote{Cf.\ \cite{Bell:2004aa}:53.} This fact shows that according to him, the general category of what exists is divided into two subsets: local and non-local beables. The former must be defined in 3-dimensional space and associated with bounded portion of spacetime; while the latter can be either defined in high-dimensional spaces or in the usual 3D space.\ For instance, he considered the state vector defined in configuration space as a proper beable in his essay \emph{Beables for quantum field theory}. Another important example is given by the multi-field interpretation of the $\psi$ field in Bohm's theory. According to this approach, given a $N$-particle system one projects the $\psi$ values  defined on configuration space into multi-field values in 3-dimensional space, i.e.\ one associates a particular field value not with individual points (as one would do with usual fields), but with $N$-tuples of points. In this manner, the multi-filed determines the motion of a configuration of particles in physical space. This new object is a beable because it is considered a real physical item living in 3-dimensional space, and it is non-local since its value is specified for a configuration of $N$ points and not for a single point.\footnote{For details on this particular perspective cf.\ \cite{Romano:2017aa}.\ Clearly, such type of beables have been admitted also by Bohm himself since he explicitly considered the wave function a real physical field, as said in the previous section. Another recent example of non-local beables can be found in \cite{Smolin:2015}.}

In sum, from our discussion we can say that Bohm and Bell's\footnote{It is worth stressing that Bell did not explicitly endorsed a metaphysical infinitism as Bohm did. Nonetheless, the similarities between their views are evident.} reflections concerning the ontology of physical theories led to the following ideas:
\begin{itemize}
\item the beables of a theory $T$ are fundamental in the context of the framework under consideration, i.e.\ they may be non-fundamental in another, deeper theory;
\item the beables must explain all the physical phenomena lying within the domain of application of $T$;
\item several ontologies can be proposed to explain the same set of physical phenomena;
\item it is possible to implement different beables within the same theoretical framework postulating a multi-category ontology;
\item beables can be non-local and defined either in 3D space or in higher-dimensional spaces;
\item finally, it is possible to implement different beables at different energy/length scales (scientific pluralism).
\end{itemize}

Since the PO programme evolved from the work of these physicists, let us then see what has been kept and what has been left behind. 

\section{The Primitive Ontology Approach}
\label{PO}

The PO approach is a normative perspective about the construction of physical theories since it provides a set of requirements that any theoretical framework should met to be considered empirically and metaphysically consistent.\ Indeed, according to the proponents of this view, well-defined theories share a common architecture deriving from these constraints.\footnote{Here I follow the modern presentation of this perspective contained in \cite{Allori:2013ab, Allori:2015}. NB: other proponents of the PO approach as for instance R.\ Tumulka, S.\ Goldstein or N.\ Zangh\`i may have different opinions concerning this programme and its main goals. For details about the common structure of PO theories see \cite{Allori:2008aa}.} 

Following Bell\footnote{For the sake of conceptual accuracy, it should be stressed that the PO approach not only is rooted in Bohm and Bell's works, but it is also inspired by Einstein, de Broglie and Schr\"odinger's reflections on and critiques to the ontology of QM. These physicists, indeed, explicitly rejected the instrumental philosophy of standard quantum theory and tried to provide an unambiguous metaphysical picture for this theoretical framework. Given the space limit, here I cannot elaborate of their influence on the PO approach. However, proponents of this perspective in several occasions---as for instance Nino Zangh\`i in many conversations and Valia Allori (private communication)---emphasize their influence.}, also in this programme the mathematical structure of a theory $T$ is divided into two subcategories.\ Firstly, there are ``primitive'' variables provided with a physical meaning;\ they represent matter and refer to real objects precisely localized and moving in 3-dimensional space (or in space-time), which is generally considered a real substance as well.\ These are the PO of the theory, the fundamental entities postulated by $T$ which constitute the building blocks of macroscopic reality.\ Remarkably, the PO defines the observable quantities of $T$---i.e.\ the properties of physical systems---and its symmetries, since it establishes which entities remain invariant under symmetry transformations.\footnote{For technical details see \cite{Allori:2008aa}. More on this later.}\
On the other hand, $T$ contains mathematical structures that are responsible for the dynamical evolution of the primitive ontology, without representing material objects---the non-primitive (or nomological) variables (cf.\ footnote 5).\ An example is given by the wave function in QM: in every PO theory $\psi$ is not considered a physical substance, but rather an essential mathematical tool useful to formulate empirically adequate laws for the dynamical history of the primitive ontology, similarly to the parameters of mass and charge.\footnote{Arguments against wave function realism can be found in \cite{Allori:2013ab, Allori:2015, Allori:2020}.}  

Furthermore, the PO has explanatory power since the macroscopic reality is metaphysically dependent on the primitive ontology of particular theories and its dynamical laws.\ This dependence relation can be intuitively characterized as mereological composition:\ the entities of a PO interact and form nuclei and atoms, which in turn aggregate into molecules and more complex biological organisms, arriving eventually to the macroscopic realm.\footnote{The exact rules of mereological composition as well as the dependence relation based on them have not yet been investigated in detail in the context of PO. This will be the subject for future research.}\ For instance, according to BM, GRWm or GRWf, the macroscopic regime is literally composed by particles, matter density fields or flashes respectively, meaning that such PO theories explain the existence of macroscopic objects in terms of the motion and interaction in space of their different mereological primitives.\ Thus, the properties and the behavior of macroscopic objects depend on---i.e.\ are reduced to---the properties and interactions of their fundamental constituents. 
Clear examples of these explanations are macroscopic measurement outcomes.\ For instance, in BM measurements of spin are explained via---and reduced to---particles' positions and their dynamical evolution as shown in \cite{Bell:2004aa} Chapter 17 and \cite{Durr:2004c}. This example  is generalizable (i) to every other PO theory and (ii) to every observable, since these frameworks provide rigorous descriptions of the physical processes taking place in measurements situations, and then also detailed explanations of the obtained results.\footnote{For technical details cf.\ \cite{Durr:2004c} and \cite{Goldstein:2012}.} All these are remarkable virtues, because observation constitutes the only connection between theory and experience. 

To conclude this brief presentation, it is worth stressing that the PO approach puts very few constraints on the selection of the primitive variables: the latter must be microscopic, since they must explain the macroscopic regime, and located in 3-dimensional space.\footnote{The reader may refer to \cite{Allori:2015} Sections 4 and 5 for arguments in favor of the requirement of ``microscopicality'' and 3-dimensionality.}\ Referring to this, indeed, Allori states that
\begin{quote}
there is no rule to determine the primitive ontology of a theory. Rather, it is just a matter of understanding how the theory was introduced, it has developed, and how its explanatory scheme works (\cite{Allori:2013ab}:65).
\end{quote} 
\noindent The definition of the primitive variables of a given theoretical framework depends on the metaphysical assumptions used to construct it, and may vary from a theory to another, as witnessed by the several PO theories available.\ 
However, the primitive ontology is never chosen \emph{a posteriori}, i.e.\ it is not read off from the mathematics of a theory. Rather, the PO itself provides an interpretation of the formalism: in the process of theory construction, a scientist selects a particular PO and will use the appropriate mathematical structures to implement her choice.
\vspace{2mm}

Let us now turn to the differences between the PO programme w.r.t.\ the previous approaches. To this regard, \cite{Allori:2013ab, Allori:2015, Allori:2020} stress two related crucial points\footnote{To my knowledge Allori is the first that pointed clearly out the differences between the PO and local beables. In literature, however, such differences are often neglected and these approaches are frequently conflated. See for instance \cite{Esfeld:2014ac} or \cite{Tumulka:2016b}.}: firstly, not every local beable is necessarily a PO. For instance, electromagnetic fields in classical electrodynamics can be interpreted as \emph{nomological} variables instead of primitive ones given their role in the theory: not only there is asymmetry between particles and fields---i.e.\ the particles can generate fields, but not vice versa---but also the latter act as mediators of particles' interactions. These facts provide indications that fields are not fundamental entities. Consequently, $\textbf{E}$ and $\textbf{H}$ are not part of the PO, i.e.\ they do not represent matter in space, but describe how particles move. Thus, they are part of the explanatory machinery of the theory. Hence, some local beables can be nomological, while PO must be exclusively material. 

Secondly, if every local beable of a given theory $T$ is considered part of the PO, then $T$ may lose symmetry properties.\ A simple example is provided by the symmetry of time reversal in electromagnetism: if $\textbf{E}$ and $\textbf{H}$ would be real physical entities, one would expect that, under time-reversal, they would still represent possible ways in which fields may be. However, the transformations $\textbf{E}(t)\rightarrow \textbf{E}(-t)$ and $\textbf{H}(t)\rightarrow \textbf{H}(-t)$ are not solutions of the Maxwell's equations. On the contrary, particles' trajectories are invariant under time reversal.\ Thus, if $\textbf{E}$ and $\textbf{H}$ are considered part of the PO, then classical electromagnetism would lose the symmetry of time-reversal. Hence, Allori concludes the PO preserves the symmetries of the theory, while sometimes local beables do not.\footnote{This is another elegant argument against the reality of fields.\ More details are given in \cite{Allori:2020} Sections 4.2 and 4.3.} 

In addition, one can also find other important differences between these perspectives. Firstly, a given PO theory $T$ defines a set of entities which are considered fundamental \emph{tout court}, since they are considered the building blocks of nature.\ According to the PO perspective, indeed, our macroscopic realm is completely ontologically reduced to the primitive variables of $T$ and its dynamical laws, which are considered the fundamental elements of reality.\footnote{To this regard, Allori writes that in the PO approach ``macroscopic objects are thought to be fundamentally composed of the microscopic entities the PO specifies. As such, the PO approach is (ontologically) reductionist, at least to the extent that it allows to make sense of claims like the PO being ``the building blocks of everything else'', and of the idea that macroscopic regularities are obtained entirely from the microscopic trajectories of the PO'' (\cite{Allori:2018}:71-72).} On the contrary, Bohm stated that a particular ontology should be valid and reliable only within the domain of application of a given theory. Similarly, Bell stressed that we may have two different sets of local beables at different energy/length scales, implying that a certain theory $T$ postulate an ontology which is \emph{theory dependent} and only relatively fundamental.

Secondly, the PO approach does not admit different ontologies at diverse scales.\footnote{If BM would be the correct description of reality at a fundamental level, we would have ontological continuity between the classical and the quantum scale. Referring to this, not only Allori is inclined to postulate particles as the fundamental PO, but also she argues that the ontological discontinuity between classical mechanics and GRW theories is an argument against the latter (cf.\ \cite{Allori:2018}).\ Clearly, I cannot speak for other proponents of this programme.}\ Indeed, if either GRWm or GRWf were the correct description of the world, the non-fundamental particle-like character of classical mechanics would emerge and explained in terms of the fundamental GRW PO. However, since we do not have direct access to the fine-grained nature of the primitive variables, we can consider the flashes/matter fields as if they were particles for explanatory purposes, recovering the appearance of classical corpuscles:
\begin{quote}
[a]t the level of microphysics we may have flashes or a continuous field, but at some mesoscopic level they produce trajectories as if they are produced by particles. So, even if the microscopic PO is not one of particles, there is a mesoscopic scale in which they behave as \emph{if} they are in the sense that from that level up to the macroscopic level the explanation is the same as if they were particles (\cite{Allori:2018}: 73).
\end{quote}

\noindent In this case, then, we would know that at the fundamental level the PO is not corpuscular, even though we would keep talking of particles for explanatory aims.\ Clearly, Bohm and Bell pluralist attitude seems to be lost in favor of a foundationalist perspective.\ Consequently, infinitism is abandoned as well. 

Thirdly, contrary to the multiple-category ontologies endorsed by Bohm and Bell, PO theories explain our manifest image of the world postulating exclusively a one-category ontology, i.e.\ the macroscopic regime is reduced to a single type of physical objects and their evolution.
However, this reduction may be \emph{non-eliminative}: for instance, as state above, in the case a GWR-type theory would postulate the correct PO, particles would be stil considered an effective description of reality from a mesoscopic to a macroscopic scale. Thus, for all explanatory purposes a classical particle ontology can be maintained. 

Finally, as we have seen in the previous sections, Bohm and Bell conceived the existence of non-local beables, whereas in the PO perspective these cannot be considered material entities. Thus, they cannot be part of the primitive ontology of a theory, and must be included within its nomological variables. In sum, if for Bell and Bohm what physically exists can be defined in terms of local and/or non-local beables, the PO can only be a subset of the local beables. Thus, the PO programme imposes stricter criteria to determine the fundamental elements of a theory w.r.t.\ the preceding approaches.

In this section we saw that the PO approach diverges from the views of its pioneers, since it employs different metaphysical assumptions and methodological principles that lead to diverse pictures of reality.\ Thus, we should be aware of these dissimilarities to not conflate these approaches. To conclude the essay, let us now consider the most recent development of the PO programme.

\section{Beyond the PO: Minimalism and Fundamental Ontology}
\label{ED}

Starting from the modern formulation of the PO programme, Esfeld and Deckert radicalized the notion of primitive ontology eliminating the theoretical dependence of the PO: ``to our mind, it is inappropriate to speak of the ontology of this or that physical theory. Ontology is about what there is'' (\cite{Esfeld:2017}:12-13). Hence, they move from the notion of ``primitive ontology of a theory $T$'' to ontology \emph{tout court}, developing their metaphysical proposal independently of any particular theoretical framework.

To achieve this result, the authors introduce a background independent atomistic ontology which seeks to recover the predictions of classical mechanics, QM and QFT via the definition of (i) different dynamical laws to be applied at the relevant energy/length scales, and (ii) appropriate typicality measures.\footnote{For technical and metaphysical details of this proposal see \cite{Esfeld:2017}, Chapter 2.} More specifically, they formulate a relationalist ontology of propertyless, finite and permanent \emph{matter points} uniquely individuated by their mutual distance relations---the only property of these points is position. The change of these relations, which is taken as a primitive fact, constitutes the dynamical aspect of the proposal. Referring to this, the authors suggest a general dynamics of matter points in terms of the first derivative of a given configuration $\Delta$ with respect to time $t$:
\begin{align*}
v_t(\Delta_t)=\frac{d}{dt}\Delta_t, \forall{t}\in\mathbb{R}.
\end{align*}
\noindent Clearly, to describe actual modifications of distance relations at various length scales, one has to provide the correct laws of motion. Here parameters as masses, charges, fields, potentials \emph{etc.} enter into the scene: they are useful mathematical tools used to implement the evolution of matter points, consequently they are not part of the PO but have an essential explanatory function.\ An explicit example is provided by the velocity field appearing in the guidance equation of BM (cf.\ \cite{Esfeld:2017}, equation 3.6), as well as those of classical mechanics and QFT (cf.\ equations 3.3 and 4.4 respectively). 

Indeed, following this metaphysical project, it is always the dynamics that changes, while the ontology of matter points remains unaltered in every theory change from the classical regime to QFT---and possibly beyond it in more fundamental theories as e.g.\ quantum gravity. Such an ontology, thus, is primitive in a new sense: it is the fundamental ontology of our physical world. These matter points are (i) ontologically independent from any other entity, and (ii) constitute the mereological basis of every object in the universe at every scale. Hence, contrary to Bohm and Bell approaches, Esfeld and Deckert's project excludes the possibility to have different ontologies at different energy/length scales. Here the PO is chosen \emph{a priori} in virtue of the guiding metaphysical principle of ontological parsimony. This ontology, in fact, is the simplest possibility to explain our macroscopic reality providing the minimal ingredients to recover spacetime and matter in motion (once the correct dynamical equations are given).\footnote{Esfeld and Deckert argue that parsimony has been a guiding principle in both philosophy and physics in their historical evolution. Indeed, given a class of theories explaining the very same set of physical phenomena, we prefer the one(s) able to explain them with the lowest number of entities and laws.} 

This proposal captures the idea of fundamentality expressed in terms of \emph{well-foundedness} (cf.\ \cite{Tahko:2018}): a dependence chain is said to be well-founded if and only if it terminates with entities which do not ontologically depend on other items.\ In the case of Esfeld and Deckert's project, matter points close the ontological dependence chain, and constitute the ultimate ingredients of reality.\ Well-foundedness, in turn, is closely related to metaphysical foundationalism, which can be defined in full generality by saying that every non-fundamental entity is dependent on some fundamental item---in this case matter points---that fully account for its being and reality. Hence, Esfeld and Deckert's project expresses the idea that reality has an ultimate foundation of matter points to which everything else is ontologically reduced---the notion of a relative fundamentality as conceived by Bohm and Bell's  is no longer maintained.\footnote{It is obvious that this project rejects Bohm's infinitism and his objections against a foundational metaphysics.}

Let me conclude this section underlying other important differences w.r.t.\ the PO programme.\ Firstly, according to Esfeld and Deckert, space-time and its geometry emerge from the configuration of matter points, contrary to the substantivalist perspective of the PO approach, where the primitive ontology of a theory is interpreted as a decoration of space.\footnote{This expression is used explicitly in \cite{Allori:2008aa}.}\ Secondly, while the PO perspective allows the postulation of several primitive ontologies, these authors accept the existence only of particles since in their opinion (i) almost all the evidence coming from the most advanced experimental research is given in corpuscular terms, and (ii) atomism can be retained not only at the classical level, but also in QM and QFT in virtue of the existence of BM and Bohmian QFTs.\ Thus, they exclude every non-particle primitive ontology. Interestingly, if the PO is a subset of the local beables, then such an atomistic ontology is a subset of the possible POs; hence, Esfeld and Deckert's project imposes stricter criteria to determine what is real w.r.t.\ the PO approach. Thirdly, these authors postulate that the number of matter points remain constant in time.\ This is an important difference w.r.t.\ the PO programme where theories with a variable number of entities have been given. For instance, the number of particles may not be constant in time, as we can see in particular Bohmian QFTs.\ This fact leads to significant differences when QFT and its phenomenology is taken into account.\footnote{In \cite{Durr:2004aa} it is postulated an ontology with a variable particle number. For a discussion of the various Bohmian QFTs with a particle ontology cf.\ \cite{Oldofredi:2018}.} 

\section{Conclusion}
\label{conc}

Starting from Bohm's infinitism and Bell's theory of local beables, we arrived to the PO programme and its most recent development, i.e.\ Esfeld and Deckert's minimalist ontology.\ Although the initial steps of this journey are in open contradiction w.r.t.\ the latest works proposed within the PO community, there is still some confusion in literature, where it is not uncommon to conflate Bell's methodological approach with the PO programme, and the latter with Esfeld and Deckert's views. Thus, in this essay I underlined the main differences among these projects. 

In particular, we saw that Bell's methodological approach shows important similarities with Bohm's ideas, sharing the belief about the provisional character of physical theories and a pluralist attitude towards their ontology. Then the PO programme has been presented emphasizing some important differences w.r.t.\ Bell's and Bohm's views. Elaborating on Allori's works, I have pointed out new divergences between these two schools of thought. Finally,  I considered Esfeld and Deckert's minimalist ontology, showing (i) the radical difference of this project w.r.t.\ Bohm and Bell's views, and (ii) its subtle dissimilarities with the tenets of the PO programme. In conclusion, I hope to have clarified at least some of the differences existing among such views, and to have shown to the reader that these approaches employ divergent metaphysical and methodological principles leading to contrasting pictures of reality.
\clearpage 
\textbf{Acknowledgements:} I warmly thank the Editor of this volume, Valia Allori, for her kind invitation to contribute to this book. I would like to thank Olga Sarno for helpful comments on previous versions of this manuscript and the reviewers for their positive and constructive feedback. This work is financially supported by the Swiss National Science Foundation (Grant No.\ 105212-175971).
\bibliographystyle{apalike}
\bibliography{PhDthesis}

\begin{thebibliography}{}

\bibitem[Allori, 2013]{Allori:2013ab}
Allori, V. (2013).
\newblock Primitive ontology and the structure of fundamental physical
  theories.
\newblock In Albert, D.~Z. and Ney, A., editors, {\em The Wave Function: Essays
  on the Metaphysics of Quantum Mechanics}, chapter~2, pages 58 -- 75. Oxford
  University Press.

\bibitem[Allori, 2015]{Allori:2015}
Allori, V. (2015).
\newblock Primitive ontology in a nutshell.
\newblock {\em International Journal of Quantum Foundations}, 1:107--122.

\bibitem[Allori, 2018]{Allori:2018}
Allori, V. (2018).
\newblock {Scientific Realism and Primitive Ontology or: The Pessimistic
  Meta-Induction and the Nature of the Wave Function}.
\newblock {\em Latosensu}, 5(1):69--76.

\bibitem[Allori, 2021]{Allori:2020}
Allori, V. (2021).
\newblock {Primitive Beable is not Local Ontology: On the Relation between
  Primitive Ontology and Local Beables}.
\newblock In Okon, E. and Romero, C., editors, {\em {Special Issue of Critica:
  The Meptaphysical Foundations of Physics}}. Universidad Nacional Aut{\'o}noma
  de M{\'e}xico.

\bibitem[Allori et~al., 2008]{Allori:2008aa}
Allori, V., Goldstein, S., Tumulka, R., and Zangh{\`\i}, N. (2008).
\newblock On the common structure of {B}ohmian mechanics and the
  {G}hirardi-{R}imini-{W}eber theory.
\newblock {\em British Journal for the Philosophy of Science}, 59(3):353--389.

\bibitem[Bell, 1987]{Bell:2004aa}
Bell, J.~S. (1987).
\newblock {\em Speakable and unspeakable in quantum mechanics}.
\newblock Cambridge University Press.

\bibitem[Bohm, 1952]{Bohm:1952}
Bohm, D. (1952).
\newblock {A suggested interpretation of the quantum theory in terms of
  ``hidden'' variables. I, II}.
\newblock {\em Physical Review}, 85(2):166--193.

\bibitem[Bohm, 1957]{Bohm:1957}
Bohm, D. (1957).
\newblock {\em {Causality and Chance in Modern Physics}}.
\newblock Routledge.

\bibitem[D{\"u}rr et~al., 2004a]{Durr:2004aa}
D{\"u}rr, D., Goldstein, S., Tumulka, R., and Zangh{\`\i}, N. (2004a).
\newblock Bohmian mechanics and quantum field theory.
\newblock {\em Physical Review Letters}, 93:090402.

\bibitem[D\"{u}rr et~al., 1992]{Durr:1992aa}
D\"{u}rr, D., Goldstein, S., and Zangh{\`\i}, N. (1992).
\newblock {Quantum equilibrium and the origin of absolute uncertainty}.
\newblock {\em Journal of Statistical Physics}, 67:843--907.

\bibitem[D{\"u}rr et~al., 2004b]{Durr:2004c}
D{\"u}rr, D., Goldstein, S., and Zangh{\`\i}, N. (2004b).
\newblock Quantum equilibrium and the role of operators as observables in
  quantum theory.
\newblock {\em Journal of Statistical Physics}, 116:959--1055.

\bibitem[Esfeld, 2014]{Esfeld:2014ac}
Esfeld, M. (2014).
\newblock The primitive ontology of quantum physics: guidelines for an
  assessment of the proposals.
\newblock {\em Studies in History and Philosophy of Modern Physics},
  47:99--106.

\bibitem[Esfeld and Deckert, 2017]{Esfeld:2017}
Esfeld, M. and Deckert, D.-A. (2017).
\newblock {\em A minimalist ontology of the natural world}.
\newblock Routledge.

\bibitem[Goldstein et~al., 2012]{Goldstein:2012}
Goldstein, S., Tumulka, R., and Zangh{\`\i}, N. (2012).
\newblock The {Q}uantum {F}ormalism and the {GRW} {F}ormalism.
\newblock {\em Journal of Statistical Physics}, 149(1):142--201.

\bibitem[Hubert and Romano, 2018]{Romano:2017aa}
Hubert, M. and Romano, D. (2018).
\newblock The wave function as a multi-field.
\newblock {\em European Journal for Philosophy of Science}, 8(3):521--537.

\bibitem[Maudlin, 2016]{Maudlin:2016}
Maudlin, T. (2016).
\newblock {Local Beables and the Foundations of Physics}.
\newblock In Bell, M. and Gao, S., editors, {\em {Quantum Nonlocality and
  Reality. 50 Years of Bell's Theorem}}, chapter~19, pages 317--330. Cambridge
  University Press.

\bibitem[Oldofredi, 2018]{Oldofredi:2018}
Oldofredi, A. (2018).
\newblock {Particles Creation and Annihilation: two Bohmian approaches}.
\newblock {\em Latosensu}, 5(1):77--85.

\bibitem[Smolin, 2015]{Smolin:2015}
Smolin, L. (2015).
\newblock {Non-local beables}.
\newblock {\em International Journal of Quantum Foundations}, 1:100--106.

\bibitem[Tahko, 2018]{Tahko:2018}
Tahko, T.~E. (2018).
\newblock {Fundamentality}.
\newblock {\em Stanford Encyclopedia of Philosophy}.

\bibitem[Tumulka, 2016]{Tumulka:2016b}
Tumulka, R. (2016).
\newblock {Paradoxes and Primitive Ontology in Collapse Theories of Quantum
  Mechanics}.
\newblock In Gao, S., editor, {\em {Collapse of the Wave Function: Models,
  Ontology, Origin, and Implications}}, pages 134--153. Cambridge University
  Press.

\bibitem[van Strien, 2019]{vanStrien:2019}
van Strien, M. (2019).
\newblock {Pluralism and anarchism in quantum physics: Paul Feyerabend's
  writings on quantum physics in relation to his general philosophy of
  science}.
\newblock {\em Studies in History and Philosophy of Science Part A}, 80:72--81.

\end{thebibliography}
\end{document}